\documentstyle[preprint,12pt,aps,prb]{revtex}
\tightenlines
\begin{document}
\newcommand{\cvo}{CaV$_{2}$O$_{5}$}
\newcommand{\mvo}{MgV$_{2}$O$_{5}$}
\title{Raman Scattering from Magnetic Excitations in the Spin Ladder 
Compounds CaV$_2$O$_5$ and MgV$_2$O$_5$ }

\author{M. J. Konstantinovi\'c$^{\ast} $}
\address{Max-Planck-Institut f\"ur
Festk\"orperforschung, Heisenbergstr.1, D-70569 Stuttgart, Germany} 
\author{Z. V. Popovi\'c}
\address{Laboratorium voor Vaste-Stoffysica en
Magnetisme, K. U. Leuven, Celestijnenlaan 200D, B-3001 Leuven, Belgium}

\author{M. Isobe and Y. Ueda} 
\address{Institute for Solid State Physics, The University of Tokio, 
7-22-1 Roppongi, Minato-ku, Tokio 106, Japan} 
\maketitle

\begin{abstract} 
We present the Raman-scattering spectra of CaV$_2$O$_5$ and MgV$_2$O$_5$. 
The magnetic contribution in the 
Raman spectra of CaV$_2$O$_5$ is found in the form of a strong asymmetric line, 
centered at $2\Delta_S$, with a tail 
on the high-energy side. Our analysis of its spectral shape shows that the 
magnetic ordering in CaV$_2$O$_5$ can be 
described using a S=1/2 two-leg ladder Heisenberg antiferromagnetic model with 
$J_{\parallel}/J_{\perp} \sim $ 0.1, and a 
small interladder exchange. The spin gap and exchange constant are estimated to 
be $\Delta_s$ $\sim$ 400 cm$^{-1}$ 
(570 K) and $J_{\perp} \sim $ 640 K. No magnon bound states are found. In 
contrast to CaV$_2$O$_5$ the existence 
of the spin-gap is not confirmed in MgV$_2$O$_5$, since we found no feature in 
the spectra which could be associated 
with the onset of the two-magnon continuum. Instead, we observe two-magnon 
excitation at 340 cm$^{-1}$, presumably
related to the top of the two-magnon brunch.  
\end{abstract}


\section{Introduction}

The spin-1/2 ladder systems, which are the intermediate situation between one-
dimensional (1D) and two-dimensional 
Heisenberg antiferromagnet (HA), show fascinating quantum effects such as 
dramatic dependence of the ground state 
on the width of the ladder (the width of the ladder is defined as a number of 
coupled chains) \cite {a11a}. Namely, 
ladders made of even number of legs have spin-liquid ground state with an 
exponential decay of the spin-spin 
correlations produced by the finite spin gap. Among them, the two-leg ladder is 
the first member of the family, with a 
largest spin gap of about 0.5 J (assuming the same coupling constant along the 
legs and the rungs). 

Realizations of the ladders are recognized in Sr-Cu-O structures
\cite {a12a},  and
recently in AV$_2$O$_5$, A=Ca, Mg 
(two-leg ladder) \cite {a1}. 
According to their crystal structures, vanadium oxides are quasi-2D layered 
materials, with spin-1/2 vanadium ions 
which form two-leg ladders coupled in the trellis lattice. The spin gap 
is observed in CaV$_2$O$_5$ to be 
around 600 K \cite {a8,a9}, and the magnetic susceptibility measurements \cite 
{a9} demonstrated almost perfect agreement with spin-1/2 two-leg ladder 
HA model. On the 
other hand, there is only contradicting evidence for the existence of the spin-
gap in MgV$_2$O$_5$ \cite {a11,a12}. 

The excitation spectrum of the ladder has been analyzed by a variety of the 
mathematical techniques. However, it was recently shown that 
the two-magnon bound state (we will use word magnon although classical magnons 
do not exist in quantum systems), 
besides usual triplet (gap mode) and two-magnon continuum, may appear in the 
excitation spectra \cite {a122}. Such a bound 
state is also predicted for the dimerized quantum spin chain \cite {a123} 
and indeed observed in CuGeO$_3$ \cite 
{a124}, using a Raman-scattering technique.
These singlet-singlet transitions (the 30 cm$^{-1}$ 
mode in the Raman spectra of 
CuGeO$_3$ is believed to be the transition between ground state and 
the two-magnon bound state which are both 
singlets) are allowed and take place via exchange scattering mechanism \cite 
{a13}. On the contrary, the Raman scattering
process on spin-gap excitations (one-magnon scattering) takes place via spin-
orbit interaction \cite {a13}, and it is 
usually negligible due to quenched orbital 
momentum in the transition metal ions.
It is therefore, desirable to study Raman scattering in magnetic ladder
materials such as CaV$_2$O$_5$ and MgV$_2$O$_5$, since these experiments
provide information about the spin-dynamics and the type of magnetic
order. In this paper, we present a study of the spin 
dynamics in CaV$_2$O$_5$ and MgV$_2$O$_5$ using Raman scattering technique.
We analyze the origin of the 
magnetic modes in the spectra and discuss the difference between magnetic 
ordering in these two compounds.

\section{Experiment}

The measurements were performed on polycrystalline samples, prepared as 
described in Ref. \cite{a1}.  As an excitation 
source we used the 514.5 nm line of an Ar$^+$   ion laser.  An average power of 5 
mW was focused with microscope
optics on the surfaces of the pellets. The temperature was controlled with a 
liquid  helium  micro-Raman cryostat 
(CryoVac, Vacuubrand GMBH).  The spectra were measured in back-scattering 
geometry using a DILOR triple 
monochromator equipped with a CCD camera. 

\section{Raman spectra of {\cvo}}

CaV$_2$O$_5$ has an orthorhombic unit cell, space group Pmmn \cite {a9}, with a 
crystalline structure formed by 
layers of VO$_5$ square pyramids. The Ca ions are situated between these layers, 
see 
Fig.1.  The pyramids are mutually 
connected via common edges and corners making characteristic V zigzag 
chains along the b axis. All vanadium 
atoms are in 4+ valence state thus carrying the spin=1/2. Effectively, this 
structure can be regarded as composed 
from spin-1/2 two-leg ladders connected with each other in the trellis lattice. 

CaV$_2$O$_5$ is isostructural with NaV$_2$O$_5$.  Therefore, the same 
irreducible representations correspond to 
the vibrational modes of CaV$_2$O$_5$  and NaV$_2$O$_5$ \cite {a14}:

$\Gamma_{opt}$=8A$_g$(aa,bb,cc)+3B$_{1g}$(ab)+8B$_{2g}$(ac)+5B$_{3g}$(bc)+ 
7B$_{1u}$({\bf
E}$||${\bf c})+4B$_{2u}$({\bf E}$||${\bf b})+7B$_{3u}$({\bf E}$||${\bf a})

The room- temperature Raman spectra of CaV$_2$O$_5$ are presented in Fig. 2. 
Comparing the phonon averaged intensities, of HH, HV and VV scattering 
configurations, with the polarized Raman scattering spectra of NaV$_2$O$_5$ 
single crystals we assign the phonon modes of CaV$_2$O$_5$. The phonon 
frequencies of CaV$_2$O$_5$ are found to be close to those in 
NaV$_2$O$_5$, as expected from similarity 
between lattice constants and interatomic distances in these two oxides.   
The low-temperature spectra  of CaV$_2$O$_5$ are shown in Fig. 3. 
In addition to phonons we found a 
strong asymmetric line at 795 
cm$^{-1}$ (1137 K) with a tail towards higher energies. This line decreases in 
intensity and broadens by
increasing the temperature, together with the intensity decrease of the 
associated continuum, see the left inset 
in Fig. 3.  Moreover, its energy is in the range of 2$\Delta_s$, estimated from 
the spin-gap measurements, thus 
suggesting its magnetic origin. 
Below, we will show by detailed discussion of the experimental results 
that this is indeed the case, i.e. that 
795  m$^{-1}$ feature corresponds to the onset of the  two-magnon continuum at 
$2\Delta_s$.

The magnetic susceptibility measurements \cite {a11} show a broad maximum around 
400 K, with a  decrease
towards zero at low temperatures, thus reflecting the low dimensionality of the 
magnetic interaction,  the singlet ground 
state and the finite spin gap.  The spin gap  is found to be 
around 660 K using electron spin resonance 
\cite {a9} and at around 500 K using nuclear magnetic resonance \cite {a8} 
experiments.  Since the magnon dispersion  
is not reported in the literature, the exchange couplings are estimated by 
comparison of the calculated and measured 
temperature dependencies of the susceptibility.  The Monte Carlo simulations 
\cite{a10,a15} showed that the magnetic 
properties of CaV$_2$O$_5$ can be indeed described  using a coupled ladder
spin-1/2 HA model (trellis 
lattice) with in-rung exchange (J$_{\perp} \sim$  600 K) five times or even an 
order of magnitude larger then the 
exchange along the legs.  The interladder exchange is found to be negligible.

Thus we start analyzing our Raman spectra using  Heisenberg Hamiltonian  for the 
ladder model:

\begin{equation}
 H=J_{\perp} \sum_{i,j-rungs} {\bf S}_i \cdot {\bf S}_j + J_{\parallel} 
\sum_{i,j-legs} {\bf S}_i \cdot {\bf S}_j 
\end{equation}

where the ${\bf S}_i$ represents a 
spin-1/2 operator at site i of the ladder. The properties of 
the magnetic excitations in the 
Heisenberg ladder have been obtained using perturbation method in the strong 
coupling 
limit \cite {a15a}. In this limit $J_{\perp} > J_{\parallel}$ the lowest
spin-triplet dispersion is \cite {a15a}:

$\omega(k)=J_{\perp}+J_{\parallel}cos(k)+\frac{3}{4}
\frac{J_{\parallel}^2}{J_{\perp}}+...$

The minimum and the maximum energies of the triplet are at k=$\pi$
($\omega \sim J_{\perp}-J_{\parallel}$) and at k=$0$
($\omega \sim J_{\perp}+J_{\parallel}$).
Thus the gap is $\Delta_s \sim J_{\perp}-J_{\parallel}$, and the total 
width of the branch $\sim 2J_{\parallel}$. Therefore, from the measurements of 
the energy onset and the width of the two-magnon continuum we could uniquely 
obtain the values of the exchange integrals $J_{\perp}$ and $J_{\parallel}$.

As we have already mentioned, the two-magnon excitations in the Raman spectra
can be observed through the exchange-scattering process. Then,  the well defined 
peak around 2$\Delta_s$ is expected in the Raman spectra of the ladder, 
due to the singularities of the one-dimensional density of (two-magnon) states 
(DOS) \cite {a15b}, or due to the appearance of the magnon bound state \cite 
{a122}. 
In the first case the peak position is exactly   2$\Delta_s$ while in the second 
case the peak can be found at  2$\Delta_s-\delta$,  $\delta$ being the 
magnon binding energy. We believe that the 795 cm$^{-1}$ line in 
CaV$_2$O$_5$ Raman spectra does not correspond to a magnon bound state due to 
the following reasons:

(1) The frustration in this material is expected to be small \cite{a15}, and the 
binding energy is then estimated to be small \cite {a122}.

(2) The intensity of the 795 cm$^{-1}$ line saturates at temperatures below 80 K. 
In the case of the bound state the linear dependence of its intensity  as a 
function of the temperature is believed to be the  fingerprint of it, without 
saturation up to very low temperatures \cite{a15d}.

Thus, we assign the onset of the magnon continuum to 2$\Delta_s$=795 cm$^{-1}$ 
and obtain the spin-gap in CaV$_2$O$_5$ to be $\Delta_s$=398 cm$^{-1}$ 
(570 K). Moreover, the width  of the continuum is estimated to be around 200 
cm$^{-1}$, see left inset of Fig. 3.
The shape of the magnon continuum in the spectra is compared with classical
noninteracting two-magnon spin-wave calculation (DOS),
$I\sim \sum_{k} \delta[E-2\omega(k)]$, and with numerical simulations 
of the spin-ladder model \cite{a15b} using exact diagonalization technique. 
These spectra are 
presented in the right inset of  Fig. 3. As expected, the DOS (full 
curve), obtained using  $J_{\perp}\sim$ 640 K and $J_{\parallel} \sim $ 70 K, 
failed to explain the nonobservation of the 
van Hove singularity, associated with the top of the two-magnon branch. It is 
well known, that in the ideal  one-dimensional systems magnons do not behave 
classically but form a continuum  of excitations \cite{a15e}.
This effect is clearly seen in exact diagonalization result (dashed curve, in 
the right inset of Fig.3) obtained with $J_{\parallel}/J_{\perp} 
\sim $ 0.1 in Ref. \cite{a15b}. However, a discrepancy between the  theory and the  
experiment is still found around 1000 cm$^{-1}$. The possible explanation may 
lay in the fact that the exact diagonalization result, \cite{a15b} has been  
obtained strictly for the two-magnon scattering process, thus neglecting higher-order magnon contributions. 
Also, we have to note, that the inclusion of the higher-order magnon processes 
would produce the scattering intensity above the two-magnon high-energy cutoff 
but it will still fail to explain the weak broad peak observed in the spectra 
around 980 
cm$^{-1}$. We suggest that such a shape of the two-magnon continuum may come 
from the next-nearest-neighbor interactions, which were not taken into 
consideration in the numerical simulations.

\section{Raman spectra of {\mvo}}

 MgV$_2$O$_5$ has an orthorhombic unit cell, space group Cmcm \cite {a16}, with 
a similar V-O layered structure as in CaV$_2$O$_5$. The structure can be 
described as a linkage of VO$_5$ pyramids, with apical oxygens along the b
axis. The V zigzag chains run along the a axis. The factor group analysis (FGA) 
\cite {a17} gives

$\Gamma_{opt}$=8A$_g$(aa,bb,cc)+5B$_{1g}$(ab)+5B$_{2g}$(ac)+6B$_{3g}$(bc)+ 
7B$_{1u}$({\bf
E}$||${\bf c})+7B$_{2u}$({\bf E}$||${\bf b})+4B$_{3u}$({\bf E}$||${\bf a})

The room temperature and the T=10 K Raman spectra of MgV$_2$O$_5$ are shown in 
Fig.4. In addition to phonon excitations, which we will not discuss here, we observe 
the continuumlike asymmetric mode around 340 cm$^{-1}$. 
This structure shows large intensity and frequency dependence as a function of 
the temperature. The temperature dependence of the spectra in the frequency 
range from 50 to 450 cm$^{-1}$ is shown in the inset of Fig. 4. The continuum 
clearly extends from 150 to 350 cm$^{-1}$ with a spectral weight cut off around 
350 cm$^{-1}$. This mode shifts to lower energies, as well, and decreases in 
intensity by increasing temperature. Such a form of the spectra and its 
temperature dependence suggests the two-magnon origin of the continuum, with a 
peak at 340 cm$^{-1}$ corresponding presumably to zone boundary magnons. 

On the other hand, no feature is found in the vicinity of 2$\Delta_s \sim$ 21 
cm$^{-1}$ (30 K) - 32 cm$^{-1}$ (45 K), \cite {a11,a12} so the existence of the 
spin gap is not confirmed by our results. However, we have to note that this 
energy range is close to the frequency limit (10 cm$^{-1}$) of our experiment. 

However, the existence of the large zone boundary magnon spectral weight in the 
Raman spectra, reflects strong deviation of the magnetic ordering in 
MgV$_2$O$_5$ from the 1D two-leg-ladder type observed in CaV$_2$O$_5$. In 
principle, such a deviation may be a consequence of the two effects

(1) There is a spin-gap in MgV$_2$O$_5$ (according to our measurements it must be 
less than 10 cm$^{-1}$) and the values of the exchange constants along legs and 
rungs are either very close to each other or the rung exchange is much smaller 
that the leg exchange (the latter case correspond to homogenous Heisenberg 
antiferromagnetic chain with a large frustration). 

(2) There is no spin-gap and magnetic ordering is quasi-2D.

For the first case, since the existence (and the value) 
of the gap is still an open question 
and since all measurements are performed 
on polycrystalline samples, it is hard to make any estimation of the exchange 
integrals. 
On the other side, if we assume 2D magnetic ordering, for example, the 
effective exchange constant can be estimated from the position of the two-magnon 
peak using relation $E_{two-magnon}=2.7 J$. In this case we obtain J=125 cm$^{-
1}$ (180 K), which is close to the energy $J_{\parallel} = $ 144 K obtained in 
Ref. \cite {a15}. However, this formula is strictly valid only for 2D square lattice 
with 
the isotropic exchange, which is clearly not the case in MgV$_2$O$_5$, even 
though the exchange couplings (along legs and rungs) are found to be all of the 
same order \cite {a15}. 
Because of that, we believe that neither one of the above 
mentioned models are applicable to MgV$_2$O$_5$, since this compound belongs to 
the class of strongly frustrated trellis lattices. Unfortunately, 
the existence of the 
spin-gap and the type of magnetic ordering in this type of lattice
are not yet known \cite {a18}.

\section{Conclusion}

In conclusion, we present the Raman-scattering spectra of CaV$_2$O$_5$ and 
MgV$_2$O$_5$. In CaV$_2$O$_5$ 
we found the onset of the two-magnon continuum at the energy 2$\Delta_s$ $\sim$ 
795 cm$^{-1}$ in the form of the 
strong asymmetric line with a tail on the high-energy side.  From the analysis 
of its spectral shape, we argue that the magnetic ordering in CaV$_2$O$_5$ can 
be described using a S=1/2 two-leg ladder Heisenberg antiferromagnetic model with 
$J_{\parallel}/J_{\perp} \sim $ 0.1, and a small interladder exchange. The spin
gap and exchange constant are estimated to be $\Delta_s$ $\sim$ 400 cm$^{-1}$ 
(570 K) and $J_{\perp} \sim $ 640 K. No magnon bound states are found. In 
contrast to CaV$_2$O$_5$ the the existence of the spin gap is not confirmed in 
MgV$_2$O$_5$. Instead, we observe a strong two-magnon excitation at about 340 
cm$^{-1}$ corresponding presumably to zone boundary magnons. These results 
reflects strong deviation of the magnetic ordering in MgV$_2$O$_5$ from the 1D 
two-leg-ladder type observed in CaV$_2$O$_5$.

{\bf Acknowledgment}

Z.V.P.  acknowledges support from the Research Council of the
K.U.  Leuven and DWTC.  The work at the K.U.  Leuven is supported by Belgian
IUAP and Flemish FWO and GOA Programs.  MJK thanks Roman Herzog - AvH for 
financial support.

$\ast$ present address: Physics Department, Simon Fraser University, 8888 
University Drive, Burnaby B.C. Canada V5A1S6.

\bibliographystyle{prsty}

\begin{thebibliography}{99}
\bibitem{a11a} E. Daggoto and T. M. Rice, Science {\bf271}, 618 (1996).
\bibitem{a12a} M. Uehara {\it et al.}, J.  Phys.  Soc.  Jpn. {\bf65}, 2764 
(1996).
\bibitem{a1} Y.  Ueda and M. Isobe, J. Magn. Magn. Mater. {\bf177}, 741 (1998).  
\bibitem{a8} H. Iwase, M. Isobe, Y. Ueda and H. Yasuoka, J.  Phys.  Soc.  Jpn. 
{\bf65}, 2397 (1996).  
\bibitem{a9} M. Onoda nad N. Nishiguchi, J. Solid. State. Chem. {\bf127}, 359 
(1996).
\bibitem{a11} M. Isobe, Y.Ueda, K. Takizawa, and T. Goto, J.  Phys.  Soc.  Jpn. 
{\bf67}, 755 (1998).
\bibitem{a12} P. Millet, C. Satto, J. Bonvoisin, B. Normand, K. Penc, M. 
Albrecht, and F. Mila, Phys.  Rev B {\bf57}, 5005 (1998).
\bibitem{a122} O.P.Sushkov and V.N.Kotov,  Phys. Rev. Lett. {\bf81}, 1941 (1998); 
K. Damle and S Sachdev, Phys. 
Rev. B {\bf57}, 8307(1998).
\bibitem{a123} G. S. Uhrig and H. J. Schulz, Phys. Rev. B {\bf54}, R9624 (1996).
\bibitem{a124} H. Kuroe, T. Sekine, M. Hase, Y. Sasago, K. Uchinokura, H. 
Kojima, I. Tanaka, and Y. Shibuya, Phys. Rev. B {\bf50}, 16468 (1994); P. H. M. 
van Loosdrecht, J. P. Boucher, G. Martinez, G. Dhalenne, and A. Revcolevschi, 
Phys. Rev. Lett. {\bf76}, 311 (1996).
\bibitem{a13} P. Fleury and R. Loudon, Phys. Rev. {\bf166}, 514 (1968).
\bibitem{a14} Z.  V.  Popovi\'c, M. J. Konstantinovi\'c, R. Gaji\'c, V. Popov, 
Y. S. Raptis, A. N. Vasil'ev, M. Isobe and Y. Ueda,  Solid State Commun. {\bf110}, 
391 (1999).
\bibitem{a10} S. Miyahara, M. Troyer, D. C. Johnston and K. Ueda,  J.  Phys.  
Soc.  Jpn. {\bf67}, 3918 (1998).
\bibitem{a15} M. A. Korotin, I. S. Elfimov, V. I. Anisimov, M. Troyer, and D. I. 
Khomskii, Phys. Rev. Lett. {\bf83}, 1387 (1999).
\bibitem{a15a} T. Barnes, E. Dagotto, J. Riera, and E. S. Swanson, Phys. Rev. B 
{\bf47}, 3196 (1993).
\bibitem{a15b} Y. Natsume, Y. Watabe and T. Suzuki,  J.  Phys.  Soc.  Jpn. 
{\bf67}, 3314 (1998).
\bibitem{a15d} P. Lemmens, M. Fischer, G. G{\"u}ntherodt, C. Gros, P. G. J. van 
Dongen, M. Weiden, W. Richter, C. Geibel, and F. Steglich, Phys. Rev. B {\bf55}, 
15076 (1997).
\bibitem{a15e} G. M{\"u}ller, H. Thomas, H. Beck and J. C. Bonner, Phys. Rev. B 
{\bf24}, 1429 (1981).
\bibitem{a16} M. Onoda and A Ohyama J. Phys; Condens. Mater. {\bf10}, 1229 (1998).
\bibitem{a17} D.  L. Rousseau, R.  P.  Bauman, and S.  P.  S.  Porto, J.  Raman 
Spectrosc. {\bf10},  253  (1981). 
\bibitem{a18} B. Normand, K. Penc, M. Albrecht, and F. Mila, Phys. Rev. B 
{\bf56}, R5736 (1997).
\end{thebibliography}

\begin{figure}
\caption
{Schematic representation of the (a) CaV$_2$O$_5$ and (b)
 MgV$_2$O$_5$ crystal structures. The thick and dashed lines represent the
 superexchange pathways.}
\label{fig1}
\end{figure} 
\begin{figure}
\caption {Raman spectra of CaV$_2$O$_5$ at room temperature in three different 
polarized 
configurations.} 
\label{fig2}
\end{figure}
\begin{figure}
\caption
{Raman spectra of CaV$_2$O$_5$ at room temperature (thick line) and at T= 10 K 
(thin line). 
Left Inset: The temperature dependence of the onset of the magnetic continuum at 
795 cm$^{-1}$.
Right Inset: The T=10 K Raman spectra compared with DOS calculation and 
exact diagonalization result for the spin-ladder model, see text.}
\label{fig3}
\end{figure} 
\begin{figure}
\caption 
{Raman spectra of MgV$_2$O$_5$ at room temperature (thick line) and at T= 10 K 
(thin line).
Inset: Raman spectra at various temperatures in the 50 to 450 cm$^{-1}$ 
frequency range.}
\label{fig3}
\end{figure} 

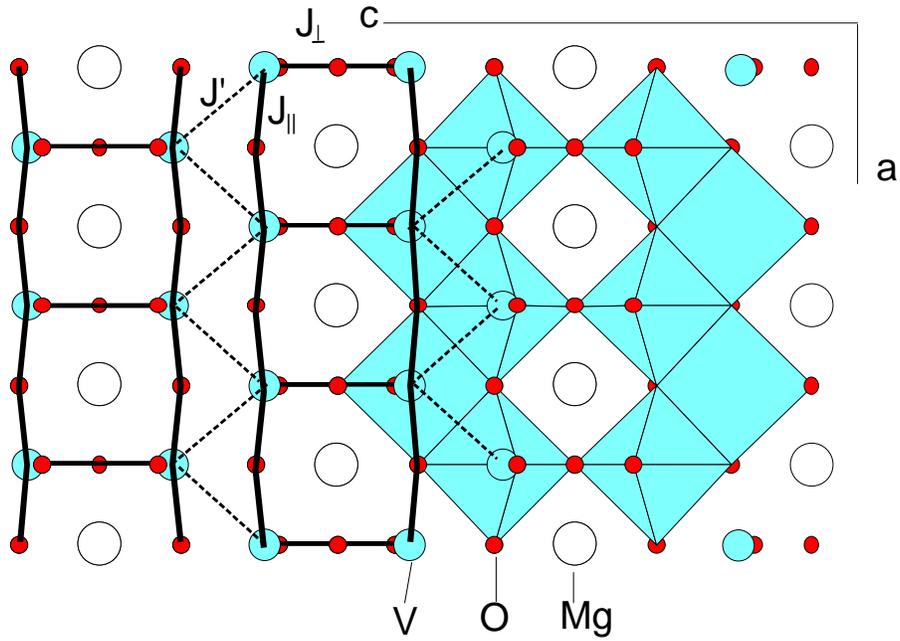
\end{document}